\documentclass[aps,prab,twocolumn,superscriptaddress]{revtex4-1}

\usepackage{amssymb}
\usepackage{lipsum}
\usepackage{amssymb}
\usepackage{bbold}
\usepackage{amssymb}
\usepackage{graphicx}
\usepackage{graphics}
\usepackage{amsmath}
\usepackage{placeins}
\usepackage{hyperref}
\usepackage[dvipsnames]{xcolor}
\usepackage{wasysym}
\usepackage{soul}
\usepackage{float}
\usepackage{array}
\usepackage{tabulary}
\usepackage{caption}
\usepackage{fancyhdr}
\usepackage{wrapfig}

\usepackage[english,ngerman]{babel}

\usepackage[normalem]{ulem}
\usepackage{graphicx}
\usepackage{color,soul}
\usepackage{subfigure}
\usepackage{dcolumn}
\usepackage{bm}
\usepackage{hyperref}
\usepackage{url}
\usepackage{multirow}
\hypersetup{
    colorlinks=true,
    linkcolor=blue,
    filecolor=blue,      
    urlcolor=blue,
    citecolor=blue,
}

\usepackage[makeroom]{cancel}

\begin{document}
\selectlanguage{english}
\title{Emittance Mapping in rf Guns}

    \author{\firstname{Benjamin} \surname{Sims}}
    \email{simsben1@msu.edu}
    \affiliation{Department of Electrical and Computer Engineering, Michigan State University, MI 48824, USA}
    \affiliation{Department of Physics and Astronomy, Michigan State University, East Lansing, MI 48824, USA}
    \affiliation{Facility for Rare Isotope Beams, Michigan State University, East Lansing, MI 48824, USA}
    \author{\firstname{Sergey V.} \surname{Baryshev}}
    \email{serbar@msu.edu}
	\affiliation{Department of Electrical and Computer Engineering, Michigan State University, MI 48824, USA}
	\affiliation{Department of Chemical Engineering and Material Science, Michigan State University, MI 48824, USA}

\begin{abstract}
    This paper discusses the trends and trade-offs between transverse \(\sigma_x\) and longitudinal \(\sigma_z\) bunch dimensions, rf injector gradient, bunch charge, and intrinsic electron mean transverse energy (MTE), where all can be chosen to be independent, and the resulting effects on emittance and transverse brightness. Using a practical example of a quarter wave normal conducting photoinjector, it is computationally found that regardless of MTE and bunch charge, there is a universal relation between the gradient $E$ and the aspect ratio of the bunch (\(\sigma_x/\sigma_z\)) leading to the highest brightness. This computational result is understood using an analytical formalism consisting of K-J Kim’s emittance formulation and a two-dimensional space charge model. The results, obtained computationally and interpreted in a robust physical framework, could therefore provide the basis for an express mapping approach for emittance forecasting when used with practical injector system design requirements and limitations.

\end{abstract}

\maketitle

\section{Introduction}\label{intro}
\begin{figure}
	\includegraphics[width=7cm]{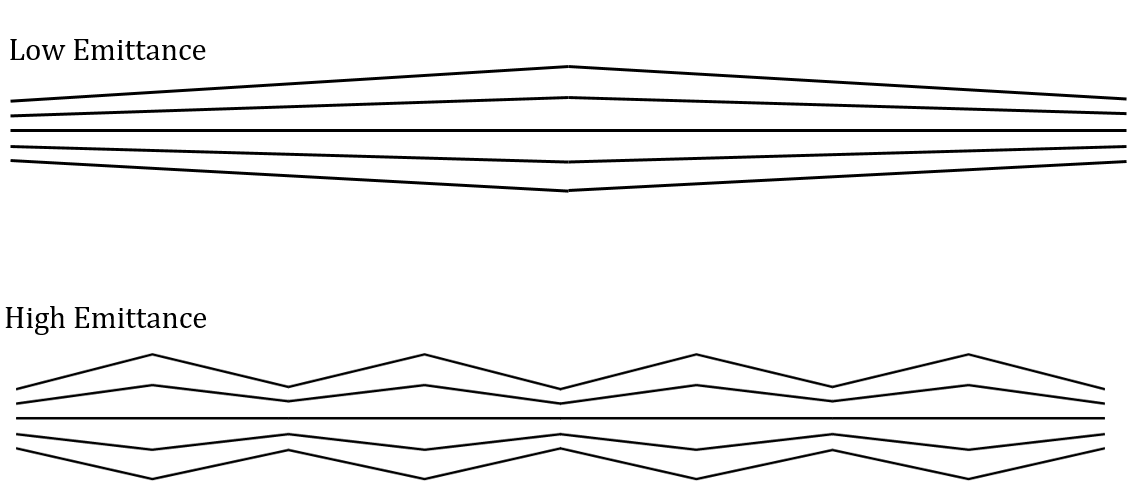}
	\caption{The concept of the low emittance gun and its effect on focusability as the beam moves down the optical lattice through focal points.}
	\label{fig1}
\end{figure}

High brightness and high current rf photo-injectors are instrumental in the progression of accelerator technology and are critical for many different applications in basic sciences, medicine, and industry \cite{nosochkov:ipac2022-tupopt046, PhysRevAccelBeams.22.023403, PhysRevSTAB.17.104401,PhysRevSTAB.15.090701}. As high brightness and high current are two counter-conflicting figures of merit
\cite{Qian:IPAC2016-THPOW020}, in order to achieve high brightness the emittance (spatiotemporal spread) needs to remain as low as possible, while maintaining a high charge. Hence, the route to achieving particular brightness/charge goals lacks universality: in each particular setting an iterative process is used to reduce the emittance \cite{PhysRevSTAB.16.073401, Papadopoulos:IPAC2014-WEPRO015,PhysRevAccelBeams.22.054602,NEVEU2023108566,Qian:IPAC2016-THPOW020, PhysRevSTAB.17.104401,PhysRevSTAB.15.100701}. Additionally, optimization is carried out along with the use of additional beam line elements like bunchers, solenoids, or beam heaters which makes the optimization process intertwined, thereby raising complexity \cite{PhysRevSTAB.16.073401, Papadopoulos:IPAC2014-WEPRO015,PhysRevAccelBeams.22.054602,NEVEU2023108566,Qian:IPAC2016-THPOW020, PhysRevSTAB.17.104401,PhysRevSTAB.15.100701}. Because the final emittance of a bunch down the linac machine can be strongly attributed to the initial beam quality, it highlights the need for an intricate understanding of initial conditions \cite{doi:10.1063/1.4913678,PhysRevSTAB.15.090701}. Therefore, an alternative approach would be to reduce complexity or, in other words, to understand the interplay between primal parameters, namely, incident laser pulse length and spot size \cite{Filippetto}. The laser parameters on the photocathode are used to engineer a desired bunch shape, thereby lowering the emittance early in the injector for a given charge, mean transverse energy, gradient and phase structure. The optimal bunch is then much simpler to maintain throughout due to a better beam focusability \cite{focusability}, as illustrated in Fig.\ref{fig1}. Practically, another benefit of this approach is a reduced number of optical lattice components necessary to achieve the desired brightness \cite{PhysRevSTAB.17.104401}, while informing optimal parameters for injector and laser design.

The described simplex approach has some interesting challenges to overcome as for the current to be high, the laser pulse dimensions need to be minimized while the charge per bunch is maximized \cite{KIM1989201}, whilst the same parameters need to be optimized to achieve the highest brightness \cite{PhysRevLett.102.104801}. The low emittance required of the bunch necessitates minimizing the space charge emittance while also managing the rf and intrinsic emittance contributions \cite{Wu:IPAC2019-WEPTS093}. The intrinsic emittance can easily be dominated by the space charge or rf emittance following emission from the cathode surface and thus low MTE cathodes are no longer a strong method of transverse emittance reduction \cite{PhysRevAccelBeams.23.070101}. In order to manipulate 
the peak current and brightness through emittance optimizations, the additional parameter of surface electric gradient has to be added. Once added, the injector design quickly becomes a multivariate optimization problem where mathematical minima for best performance are sought after using computational algorithms \cite{PhysRevSTAB.8.034202, PhysRevSTAB.16.010101}. Thanks to the simplicity of the injector emittance minimization approach, the optimization algorithm can be cross-verified analytically, thereby helping establish more universal relationships between laser and rf injector cavity subsystems, and the resulting bunch parametrization and emittance mapping.

\section{Case study setup}\label{setup}

To illustrate the described ideas, emittance of the Argonne Cathode Teststand (ACT) photo-injector was mapped and optimized. The ACT gun is a canonical quarter-wave normal conducting L-band (1.3 GHz) injector that was thoroughly described and verified computationally and experimentally \cite{ACT2019,ACTmainref}. The laser spot was defined as a radially uniform circle while the pulse length was a Gaussian distribution clipped at $3\sigma$ (99.7-rule). The ACT has two solenoids: one, wrapped around the gun, is used for emittance compensation, and one, at the end of the injector, is used for focusing the bunch onto the target. Because the main idea of this work is to optimize for minimal emittance without the use of additional focusing apparatuses, both solenoids were disabled in the simulations to isolate the fundamental emittance effects (MTE, space charge and rf) on the bunch. The simplicity of the ACT injector allows for extensive analysis and effective demonstration of the emittances at play. Together, the gradient $E$ and its phase structure, \(\sigma_x\) and \(\sigma_z\), form a parameter space that needs to be analyzed to provide the best injector and drive laser beam settings to increase the brightness. The choice of metal (low QE, short response time) versus semiconductor (high QE, long response time) photocathodes and their MTE determine the limitations of the primary laser pulse length \(\sigma_z\), and the spot/source size \(\sigma_x\). This work explores emittance optimization at 10, 30, 50, 70, and 90 MV/m to cover the full operating range of the ACT \cite{A-class} as well as illustrate the differences that each gradient requires. The contribution to the emittance from the cathode material also cannot be overlooked, and $\sim$0 meV MTE material (at Boltzmann tail operation) and a 200 meV (common approximation) MTE were compared. These examples allow us to emphasize important nuances of minimizing emittance in modern low-gradient SRF guns \cite{osti_1029479} against high and ultrahigh-gradient copper guns operated at room and cryogenic temperatures \cite{BDR}.

\

\section{Methods and Definitions}\label{methods}

The simulations were performed using General Particle Tracer (GPT) \cite{gpt}, a robust simulation program of choice in photoinjector community that has emphasis on space charge dynamics. Emittance heatmaps were generated which mapped the topography of the pulse length and spot size environment. All particle simulations were done in GPT with space charge effects enabled. Each simulation consisted of either 10 pC or 100 pC per laser pulse. These two levels of charge are considered state-of-the-art for future systems and applications, e.g. these levels are considered as main operating conditions for the low emittance injector (LEI), a core upgrade step for the LCLS-II-HE project \cite{nosochkov:ipac2022-tupopt046,Qian:IPAC2016-THPOW020}. Considered charge was spread over 1,000 macro-particles to reduce computational time when generating heatmaps. A field map of the ACT was used, and all simulations were performed until the end of the injector, 8 cm away from the cathode surface set as the origin \cite{ACT2019, A-class}. The two distinctly different MTE, 0 versus 200 meV were chosen. 0 meV was approximated as 1 meV in the simulations and represents copper cathode operated in cryogenic copper injectors \cite{PhysRevLett.125.054801}. While 200 meV represents the standard MTE expectation for most photocathodes at room temperature with emission above the threshold, i.e. outside Boltzmann tail regime; one example being Cs$_3$Sb \cite{doi:10.1063/1.3652758} which is being considered for the low emittance injector for LCLS-II HE.

The pulse length was scanned from a 1 ps gaussian pulse clipped at $3\sigma$ to a 20 ps gaussian pulse also clipped at $3\sigma$. The minimum pulse length was chosen as a cut-off to avoid space charge locking after observing the beam tracking, where not all of the particles were emitted from the surface. Space charge locking occurs when enough charge is present near the cathode surface, thereby negating the electric field on the cathode. Once this occurs, particles are still emitted from the cathode due to the photoelectric effect but are no longer accelerated by the field. This results in a reduced charge yield from the given laser pulse and a misshapen bunch dependent on when the charge locking occurred. There is an additional difficulty in GPT where the particles can’t be reabsorbed by the cathode and instead are accelerated backwards in the simulation. In many cases this breaks the simulation and the space charge locking effect needs to be avoided in GPT simulations. The spot size radius was examined from 0.4 mm to 1 mm. The minimum size was chosen again using the same criterion to avoid space charge locking. These simulation parameters were then repeated at 10, 30, 50, 70, and 90 MV/m.

\subsection{Normalized Transverse Emittance}
The transverse emittance of a bunch is defined as the area of the ellipse in the momentum phase space that the bunch occupies.
The emittance can be calculated using a statistical root mean square (RMS) approach. If the bunch is assumed radially uniform, the \(\varepsilon_x\) and \(\varepsilon_y\) emittances are identical, only the \(\varepsilon_x\) needs to be calculated. The statistical definition of \(\varepsilon_x\) is shown in Eq.\ref{eq:1} and was used to calculate all emittances that follow.

 \begin{gather}
    \varepsilon_x=\sqrt{ \langle x^2 \rangle \langle x'^{2}  \rangle - \langle x^2 \cdot x'^{2} \rangle } \label{eq:1}
\end{gather}
 
The particle locations, $x$, as well as the normalized transverse momentum, $x’$, are required to calculate the emittance. These values were computed in GPT and exported for calculation separately instead of using GPT’s built-in emittance calculation routines. GPT calculates normalized transverse emittance in the velocity space using only $\beta_x$ \cite{gpt}. Instead the normalized transverse emittance was calculated using Eqs.\ref{eq:2}-\ref{eq:6}. It was calculated as

\begin{gather}
    \beta_x = \frac{v_x}{c}     \label{eq:2} \\
    p_x^N = \frac{p_x}{mc} = \gamma \beta_x = x'  \label{eq:3} \\
    \varepsilon_x=\sqrt{ \langle x^2 \rangle \langle (\gamma \beta_x)^2  \rangle - \langle x^2 \cdot (\gamma \beta_x)^2 \rangle } \label{eq:4}
\end{gather}

To normalize the \(\varepsilon_x\) it was scaled by \(\beta_z\) and \(\gamma\) as shown in Eq.\ref{eq:6}. This was done to ensure equitable comparison regardless of the energy of the bunch resulting from other injectors or field strengths.

\begin{gather} 
    \beta_z = \frac{v_z}{c} \label{eq:5}  \\
    \varepsilon_n=\gamma \beta_z \varepsilon_x  \label{eq:6}
\end{gather}

\subsection{Brightness}
The brightness of a beam has two components, the current and the emittance. From Eq.\ref{eq:7}, the brightness is linearly proportional to current but inversely proportional to the square of the emittance

 \begin{gather}
    B = \frac{2I}{\varepsilon^2} \label{eq:7}
\end{gather}

\section{Results}\label{GPT}

\begin{figure}[]
    \includegraphics[width=0.45\textwidth, height=0.245\textwidth]{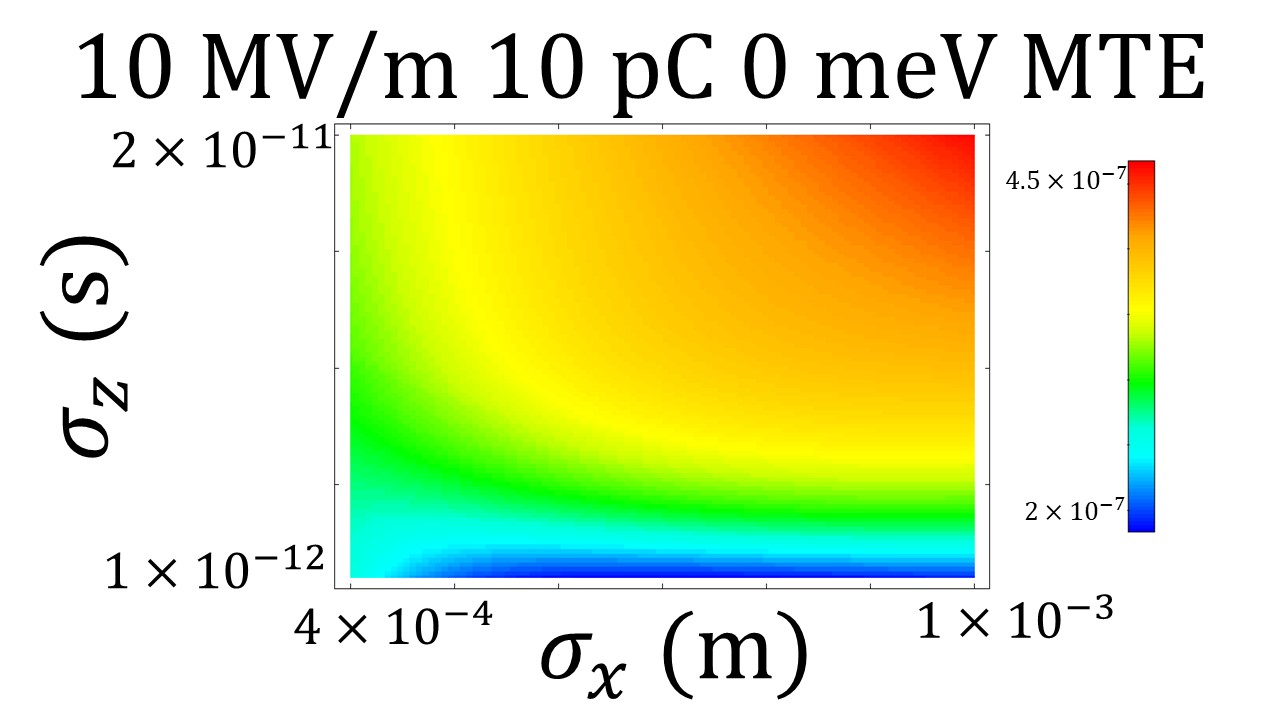}
	\includegraphics[width=0.45\textwidth, height=0.245\textwidth]{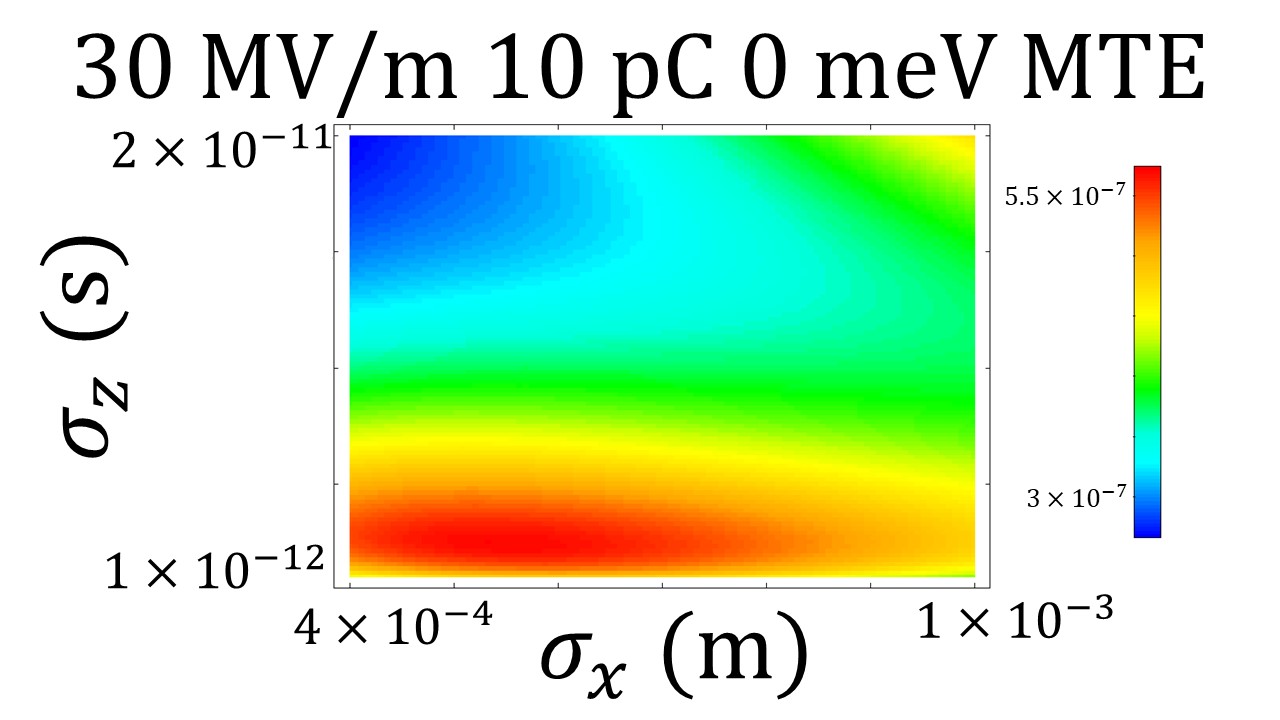}
	\includegraphics[width=0.45\textwidth, height=0.245\textwidth]{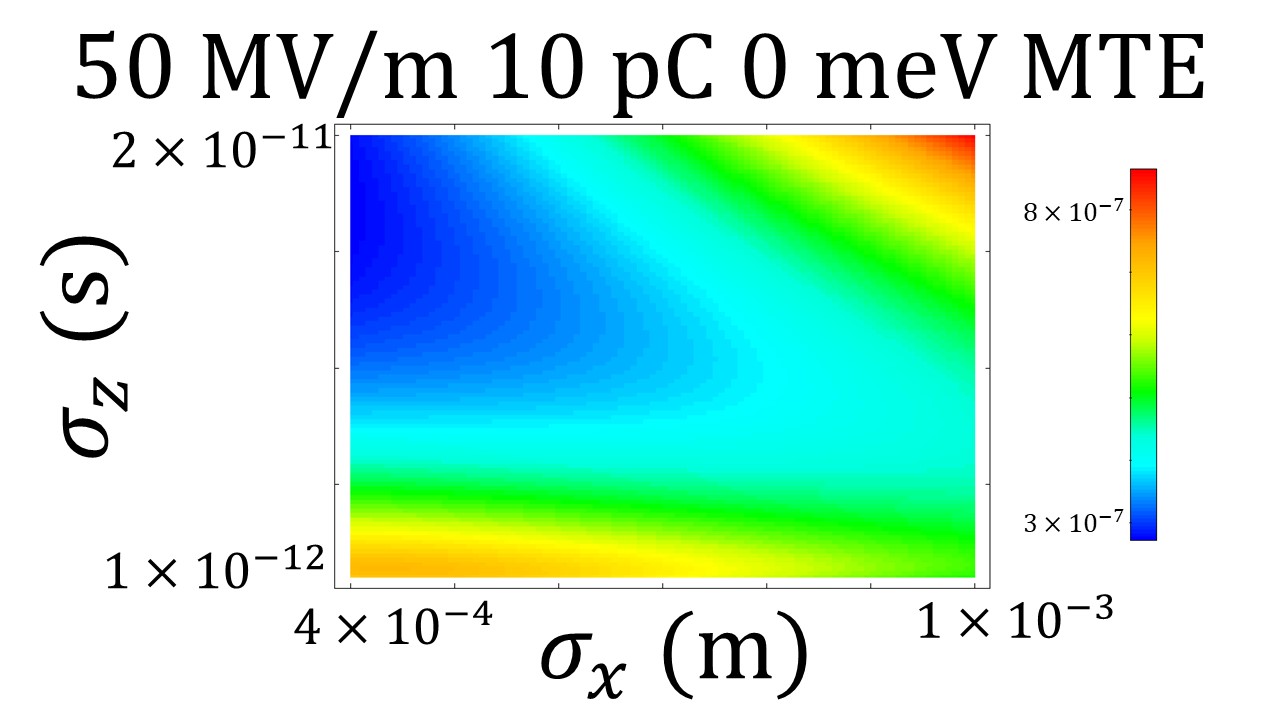}
	\includegraphics[width=0.45\textwidth, height=0.245\textwidth]{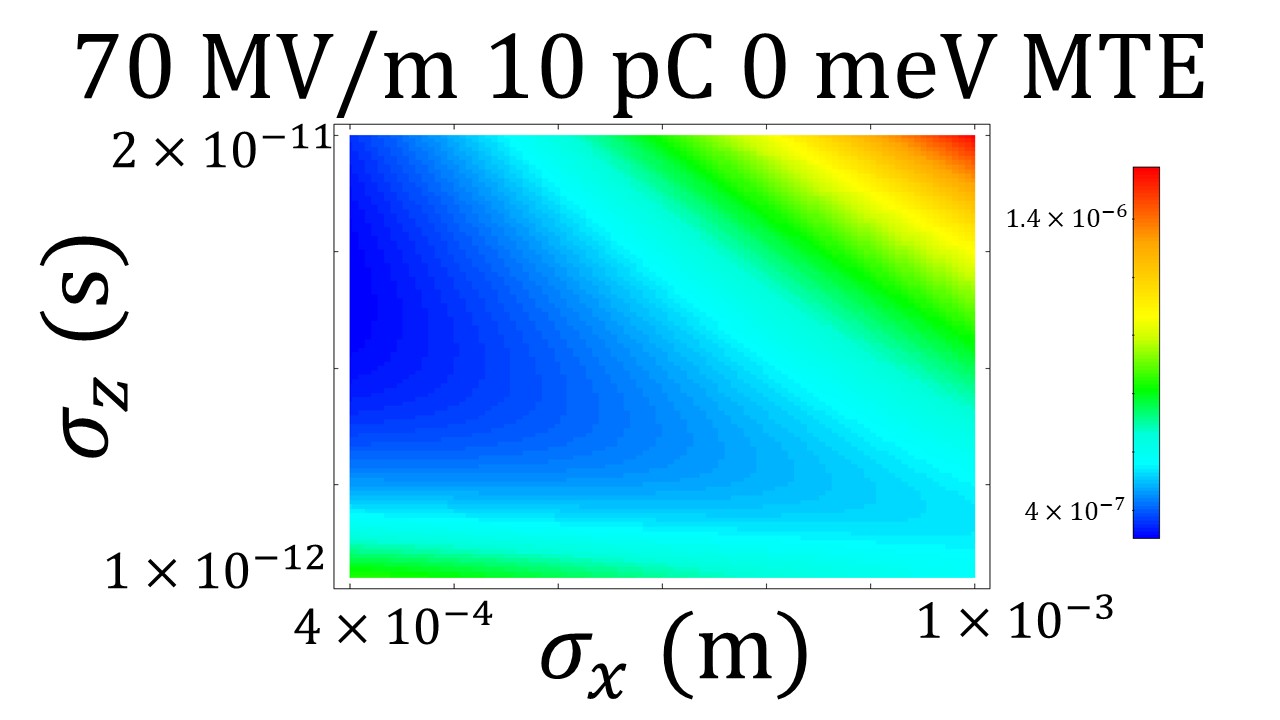}
	\includegraphics[width=0.45\textwidth, height=0.245\textwidth]{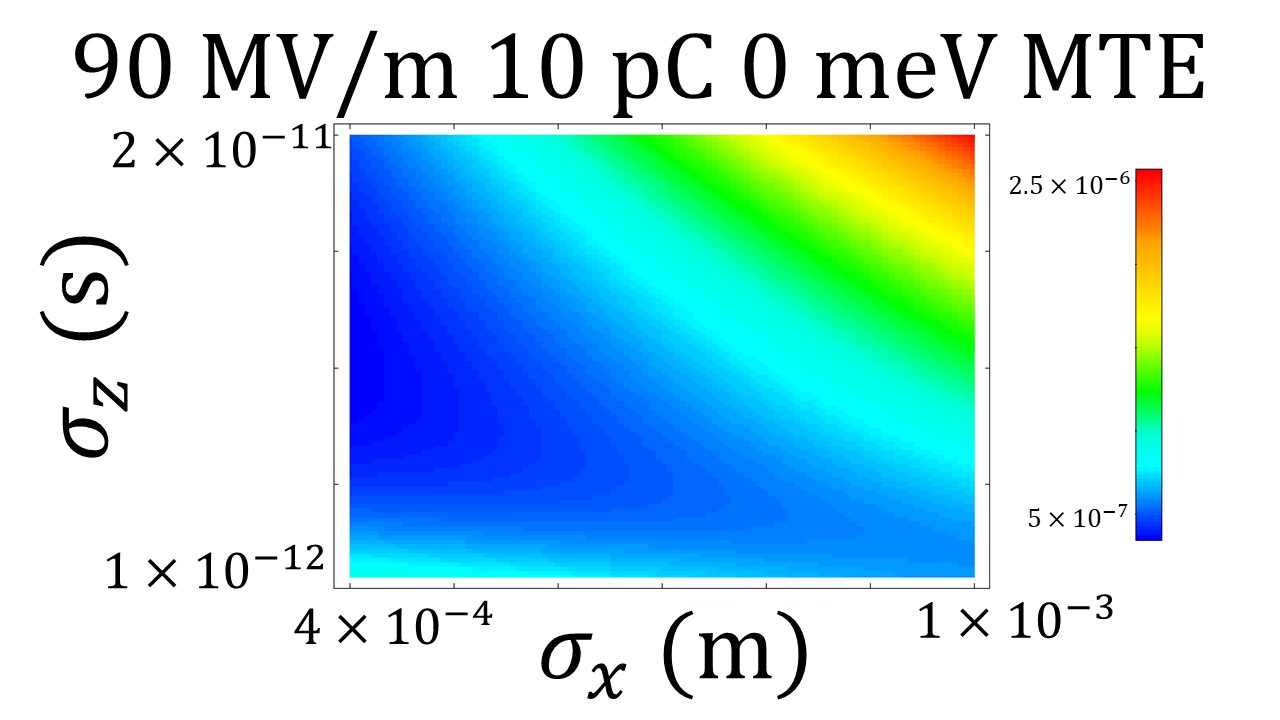}
\caption{Series of heatmaps for 10 pC and 0 MTE case study demonstrating beam form factor evolution at different gradients. Slight increase of the emittance floor as the gradient is swept from 10 to 90 MV/m is because of the rf emittance contribution due to long bunch.}
\label{Fig 2}
\end{figure}

The arrays of the 100 cell by 100 cell heatmaps presented below examine emittance in the parameter space defined by beam form factors at gradients of 10, 30, 50, 70, 90 MV/m for the described ACT photoinjector. The number of macro-particles used in the GPT simulations was set to 1,000. This was necessary for achieving high resolution of the emittance heatmaps (within a reasonable computational time frame of approximately 24 hours per heatmap), and, in turn, allowing for identifying critical trends. The cost of reducing the number of macroparticles was decreased accuracy of the individual emittance values calculated. It was estimated that the emittance was within \(\pm\)17\% of low resolution heatmap simulations conducted with 1,000,000 particles. Hence the established value of decreased resolution does not affect the basic conclusions and findings deduced from the high resolution heatmap analyses. 
\begin{figure}[]
    \includegraphics[width=0.45\textwidth, height=0.245\textwidth]{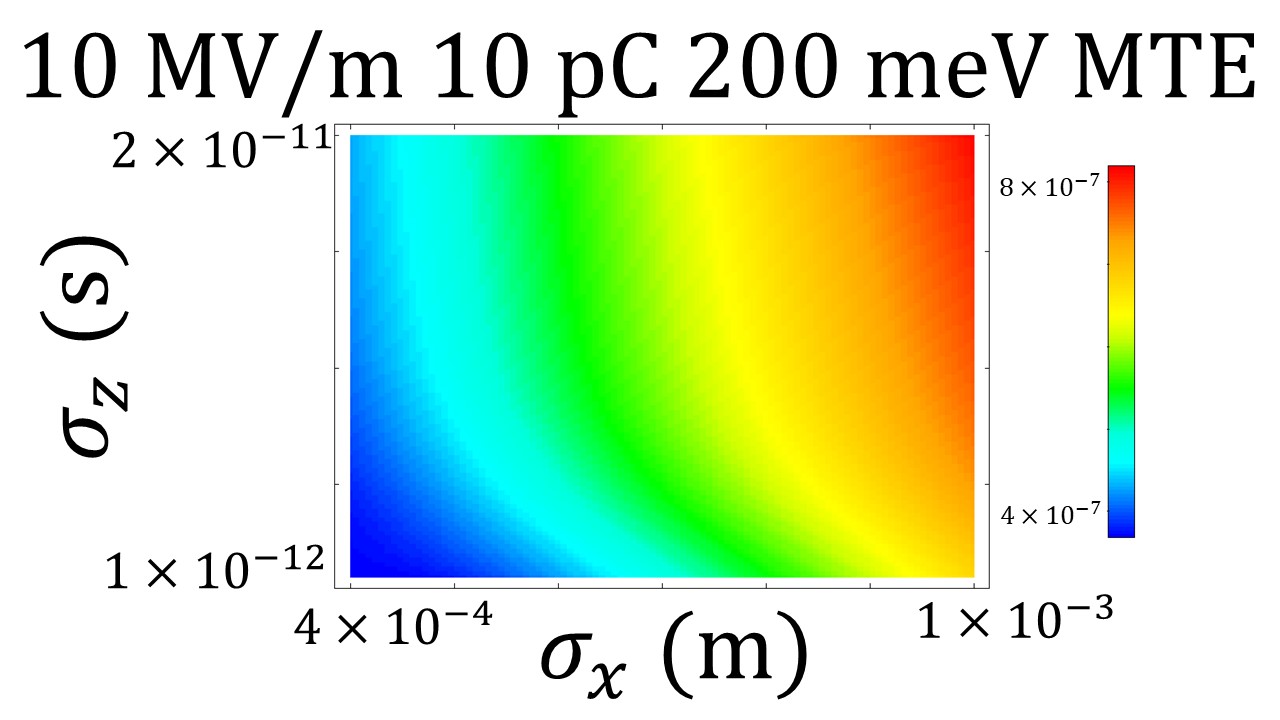}
	\includegraphics[width=0.45\textwidth, height=0.245\textwidth]{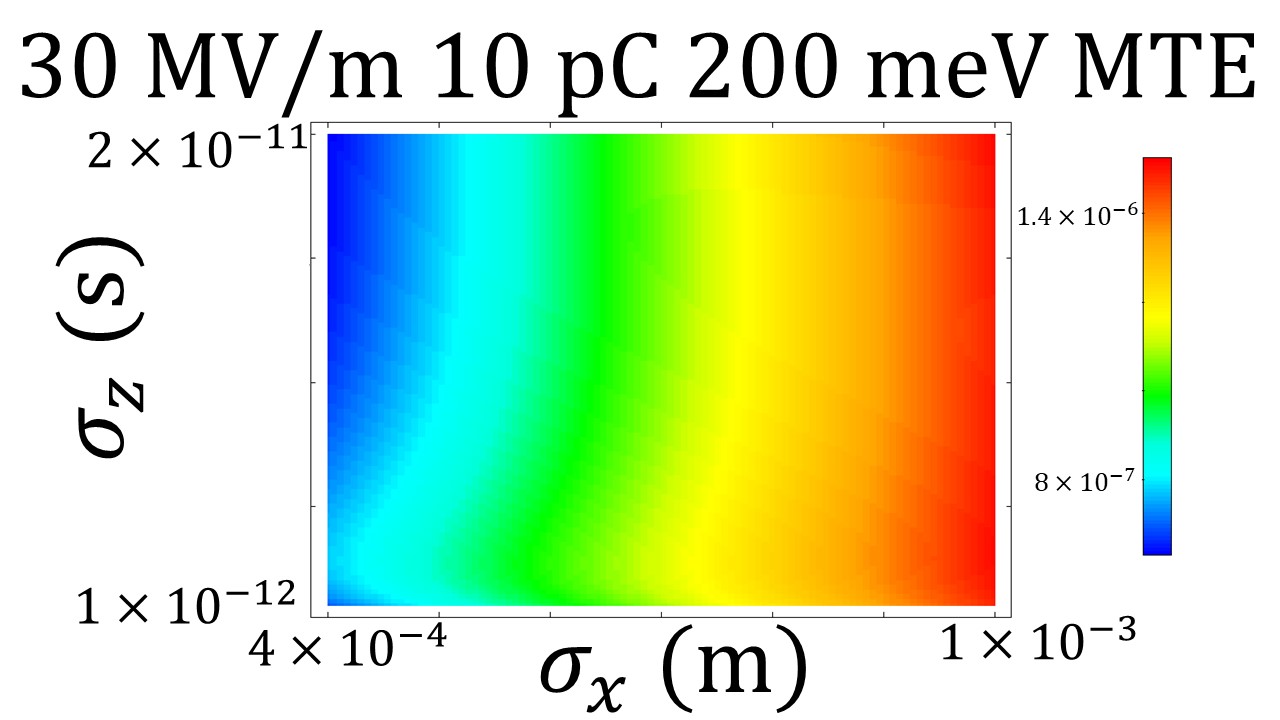}
	\includegraphics[width=0.45\textwidth, height=0.245\textwidth]{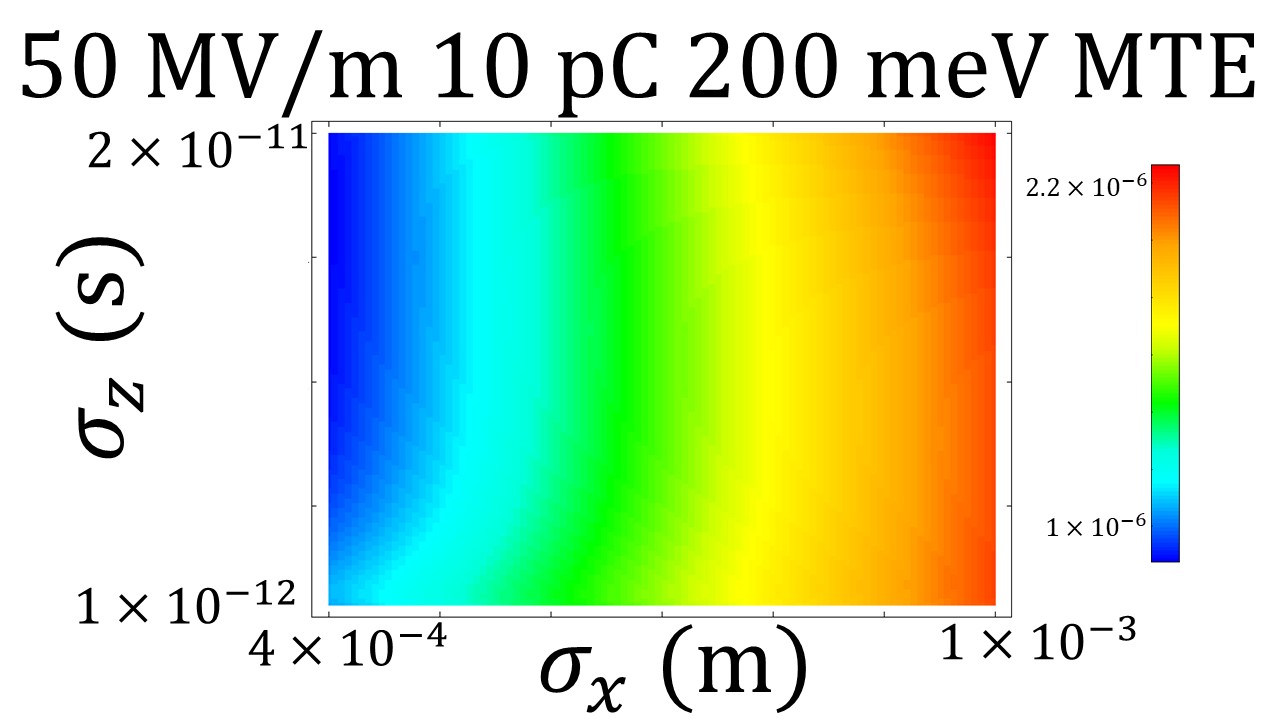}
	\includegraphics[width=0.45\textwidth, height=0.245\textwidth]{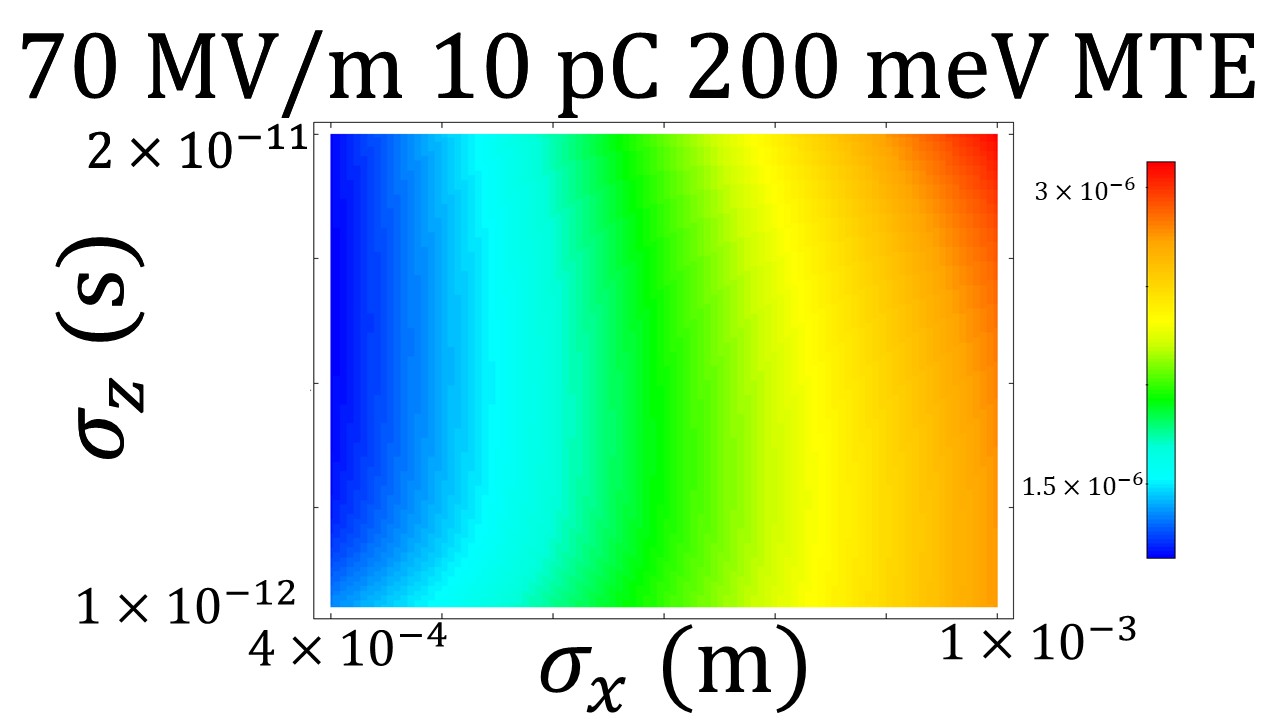}
	\includegraphics[width=0.45\textwidth, height=0.245\textwidth]{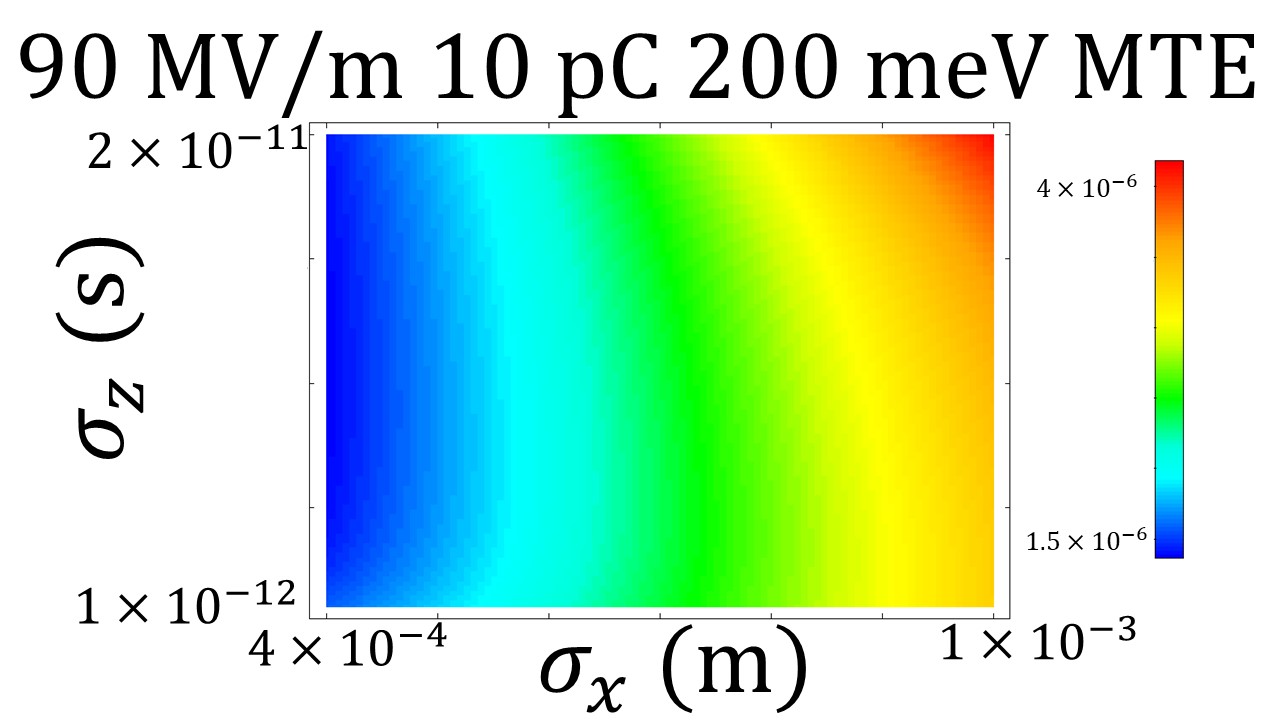}
\caption{Series of heatmaps for 10 pC and 200 meV MTE case study demonstrating beam form factor evolution at different gradients. 10 MV/m minimal emittance location shifted to a short ellipse bunch shape. Higher gradients, again, indicate cigar beam as optimal.}
\label{Fig 3}
\end{figure}

\subsection{10 pC, 0 MTE}

Fig.\ref{Fig 2} illustrates that the optimal bunch configuration is extremely dependent on the gradient, and guides on how the charge needs to be distributed to mitigate the space charge effect. More specifically, the 10 MV/m heatmap suggests the pancake regime, thereby implicating that the best emittance is achieved through the shortest pulse (smallest \(\sigma_z\)) and largest spot size (largest \(\sigma_x\)). Moving on to the 30 MV/m case study, the situation flips, and the cigar regime is implicated (blue upper left corner of the heatmap), thereby requiring much smaller \(\sigma_x\) and much larger \(\sigma_z\). When analyzed in the context of brightness, there are additional factors to consider. The brightness depends on current and the current increases as pulse length decreases. As a result, having a minimal pulse length and minimal emittance is important. The minimum emittance for each heatmap also increases with the gradient. This is due to the increasing effects of the rf emittance on the bunch as the gradient increases.

\subsection{10 pC, 200 meV MTE}

Moving from the previous, rather idealized, case to one where the intrinsic emittance is no longer effectively a zero, this produces a new set of data that differ from the previous case study. Here, the intrinsic emittance is calculated as
 \begin{gather}
    \varepsilon_{int}=\frac{\sigma_{xi}}{m_e c}\sqrt{2 \cdot m_e \cdot e \cdot MTE },\label{eq:8}
\end{gather}
 where the intrinsic emittance now depends on the initial spot size of the bunch $\sigma_{xi}$. It should, again, be noted that the resulting emittances plotted in all of the heatmaps are calculated purely based on the statistics of the energy and position of the electrons, see Eq.\ref{eq:4}.

In the 10 MV/m simulation, the lowest emittance is attained when \(\sigma_x\) and \(\sigma_z\) are on par in terms of the size, thereby requiring that the beam is no longer a pancake form-factor but rather a short ellipse. This result is expected as emittance is proportional to \(\sigma_x\) while being affected by an MTE that is now 200 times larger (1 versus 200 meV), thereby requiring \(\sigma_x\) to reduce. Otherwise, heatmaps in Fig.\ref{Fig 3} corresponding to 30, 50, 70, and 90 MV/m demonstrate the trends similar to those in Fig.\ref{Fig 2}. The higher the gradient, the more the bunch is required to shape into a cigar form factor. However, with the increased effect of the intrinsic emittance, the area of minimal emittance is much more restricted as the bunch can no longer benefit from an increased spot size. Which results in the minimum emittance staying near the smallest spot size possible.

\subsection{100 pC, 0 MTE}
\begin{figure}[]
    \includegraphics[width=9cm]{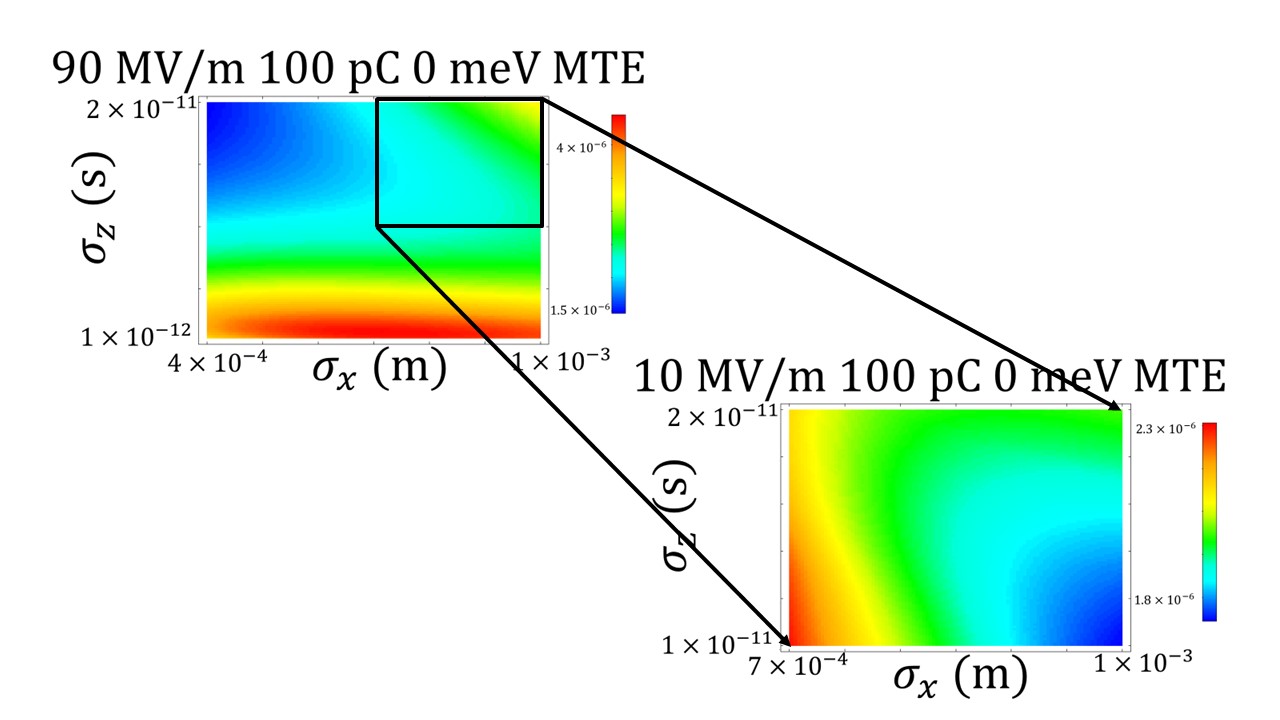}
	\caption{Demonstration of the parameter space cutout for 100 pC 0 MTE case study that had to be performed because of enhanced space charge screening effect at the cathode surface.}
\label{Fig 4}
\end{figure}

When working with the ambitiously high bunch charge of 100 pC, the $\sigma_x-\sigma_z$ range had to be adjusted for the 10 MV/m gradient case. Due to a ten-fold charge increase, it was found that no particles would leave the cathode surface due to space charge locking. The electrons would effectively create a wall of charge negating the field of the gun at the cathode, leading to electrons emitted later in the pulse returning to the cathode. The $\sigma_x-\sigma_z$ range where charge locking was not present is shown in Fig.\ref{Fig 4}. As can be seen, the maximum pulse length was held constant while the minimum pulse length boundary was increased to $10^{-11}$ s, while $\sigma_x$ was only run from 0.7 to 1 mm. The main result was that the minimal emittance location is in the lower right corner, a perfect pancake bunch.

\begin{figure}[]
    \includegraphics[width=0.45\textwidth, height=0.245\textwidth]{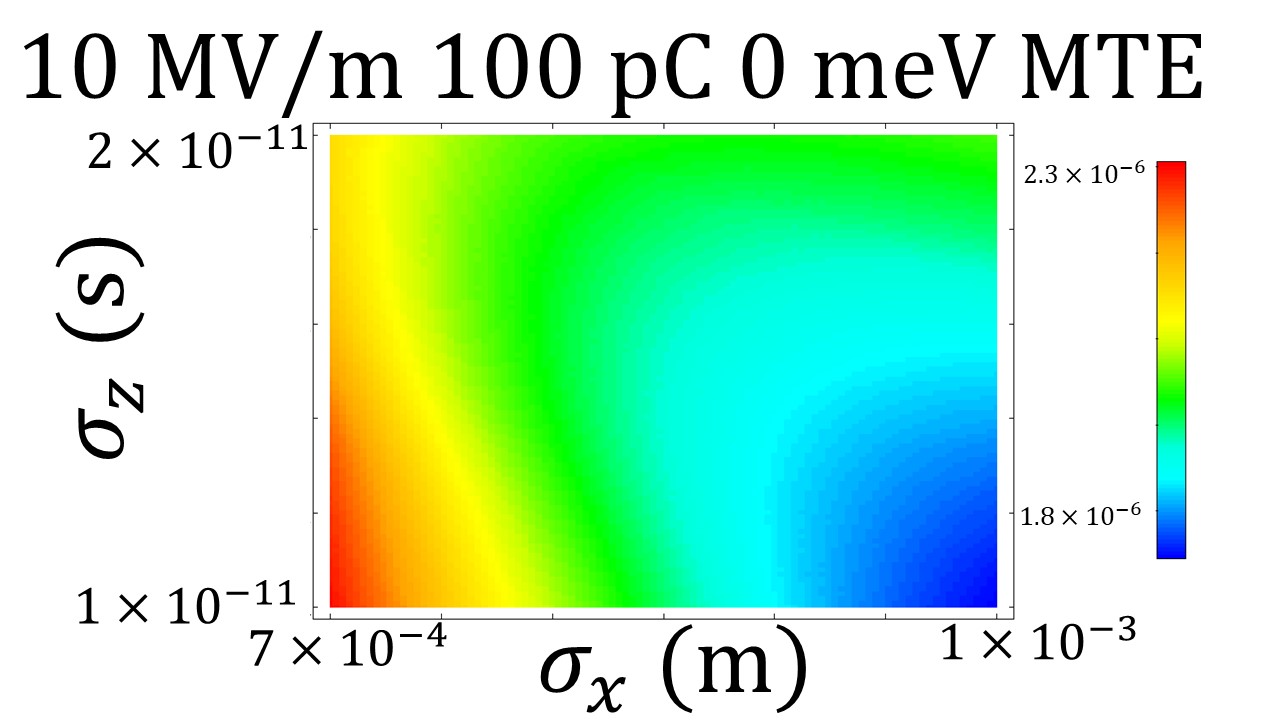}
	\includegraphics[width=0.45\textwidth, height=0.245\textwidth]{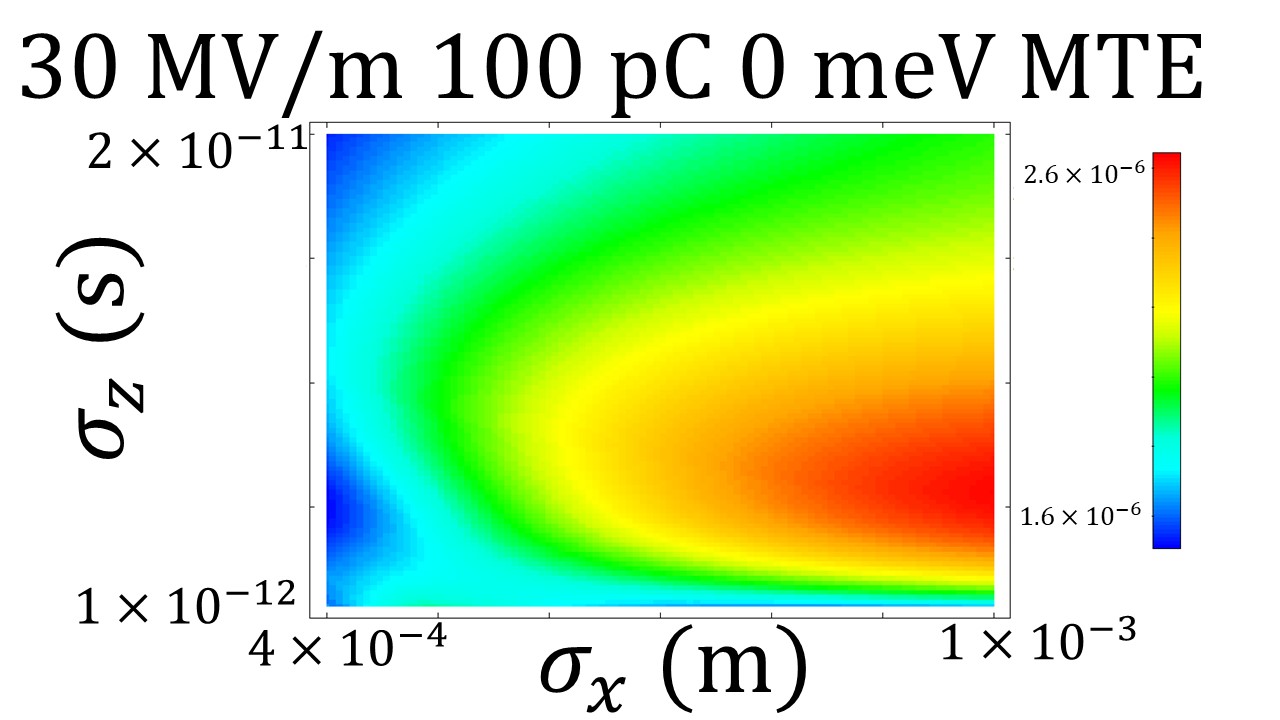}
	\includegraphics[width=0.45\textwidth, height=0.245\textwidth]{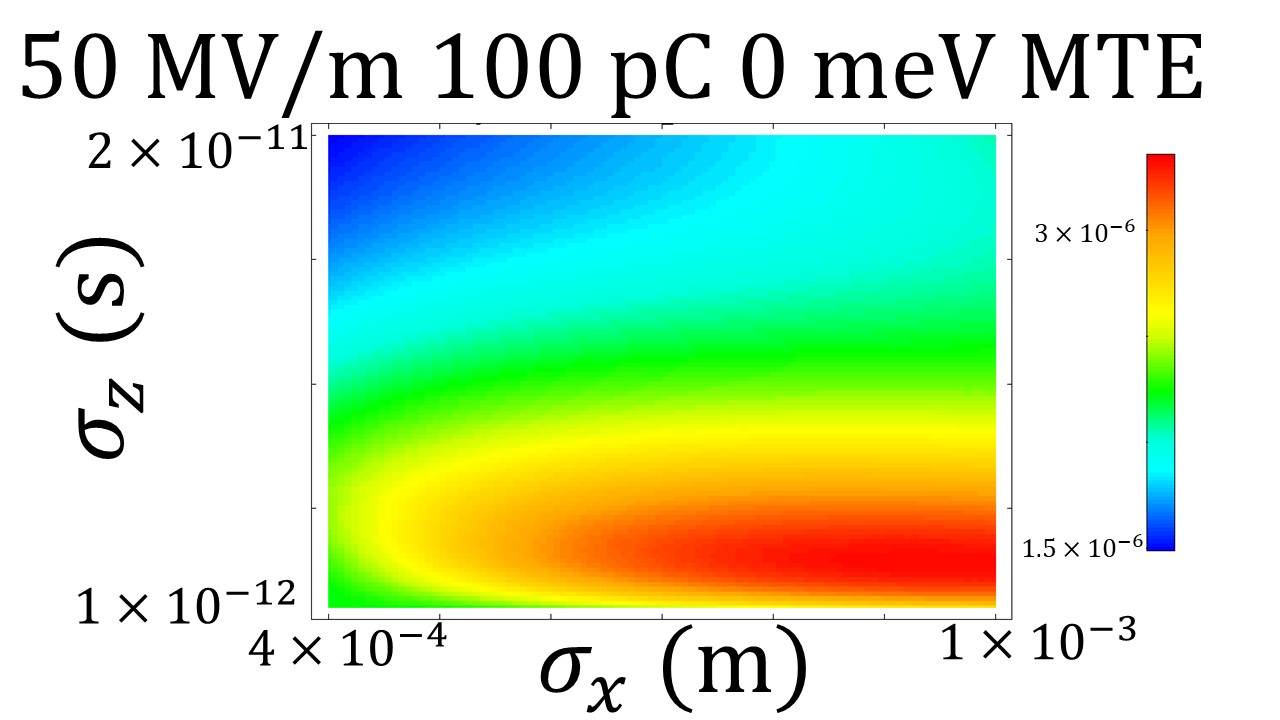}
	\includegraphics[width=0.45\textwidth, height=0.245\textwidth]{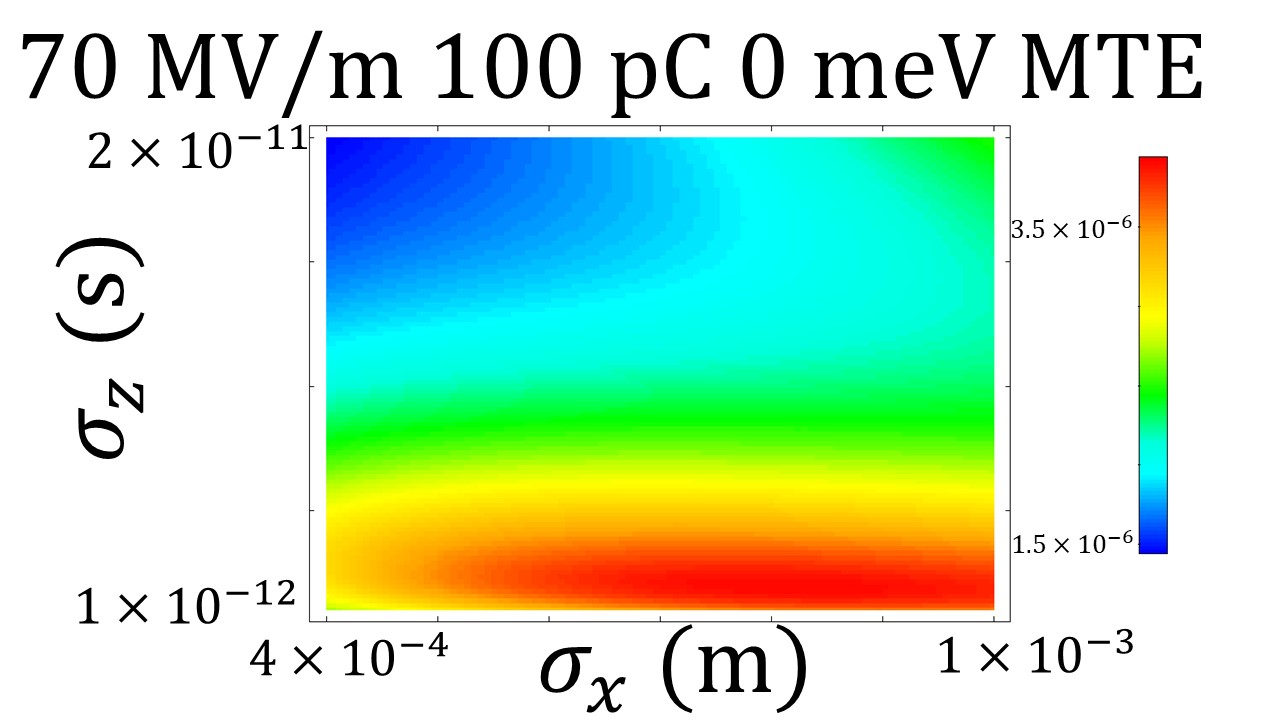}
	\includegraphics[width=0.45\textwidth, height=0.245\textwidth]{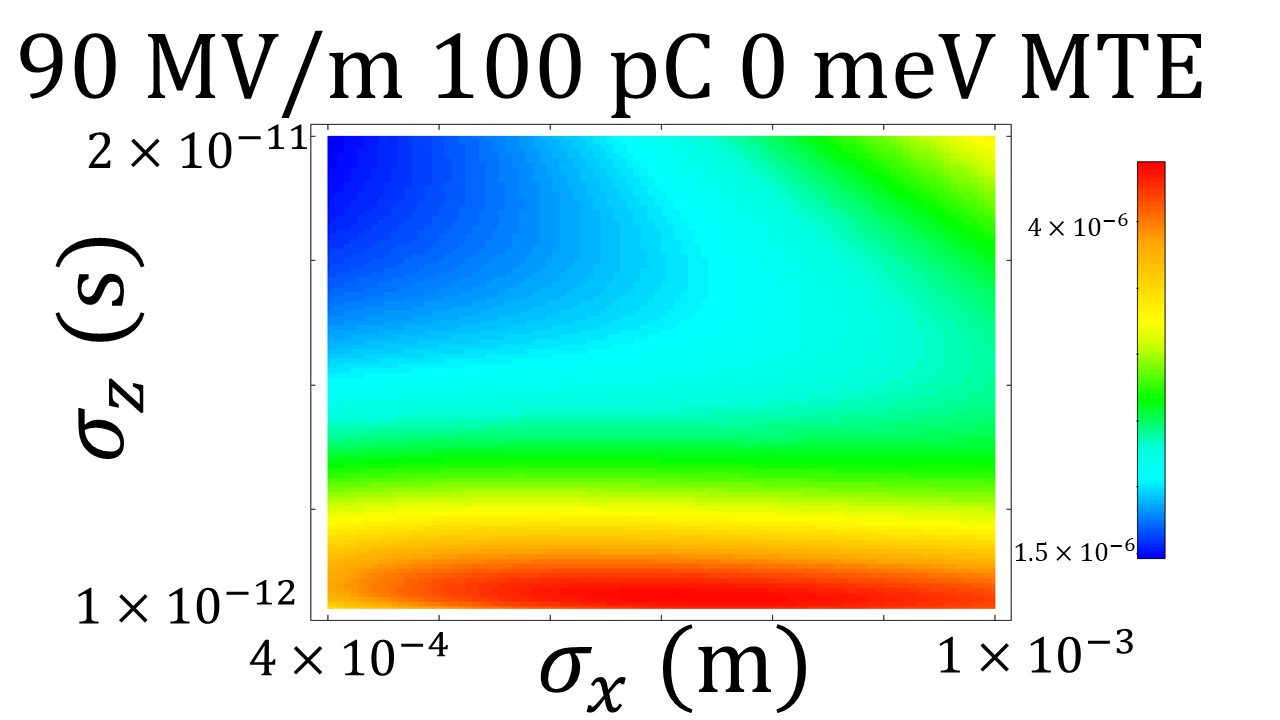}
\caption{Series of heatmaps for 100 pC and 0 MTE case study demonstrating beam form factor evolution at different gradients. The optimal emittance minima locations are identified to be in the corners indicating a strong need for either pancake or cigar bunches.}
\label{Fig 5}
\end{figure}

At 30 MV/m, it flips to opposite upper left corner, perfect cigar bunch. Expanding this result, an ideal high charge bunch at low gradients would be a perfect (infinitely thin) disk, whereas an ideal high charge bunch at high gradients would look like a line of charge. These bunch designs would be impossible with current technologies but do provide useful reference points when thinking about practical accelerators and their limitations, as well as for the designing bunch shape for a specific experiment. As the gradient grows from 30 to 90 MV/m, the range enabling minimal emittance extends, which is simply because higher gradients more strongly compensate the coulombic repulsion force. Otherwise, the emittance baseline increased because of the stronger space charge effect as compared to 10 pC bunch charge. All of these results are summarized in Fig.\ref{Fig 5}.

\subsection{100 pC, 200 meV MTE}
\begin{figure}[]
    \includegraphics[width=0.45\textwidth, height=0.245\textwidth]{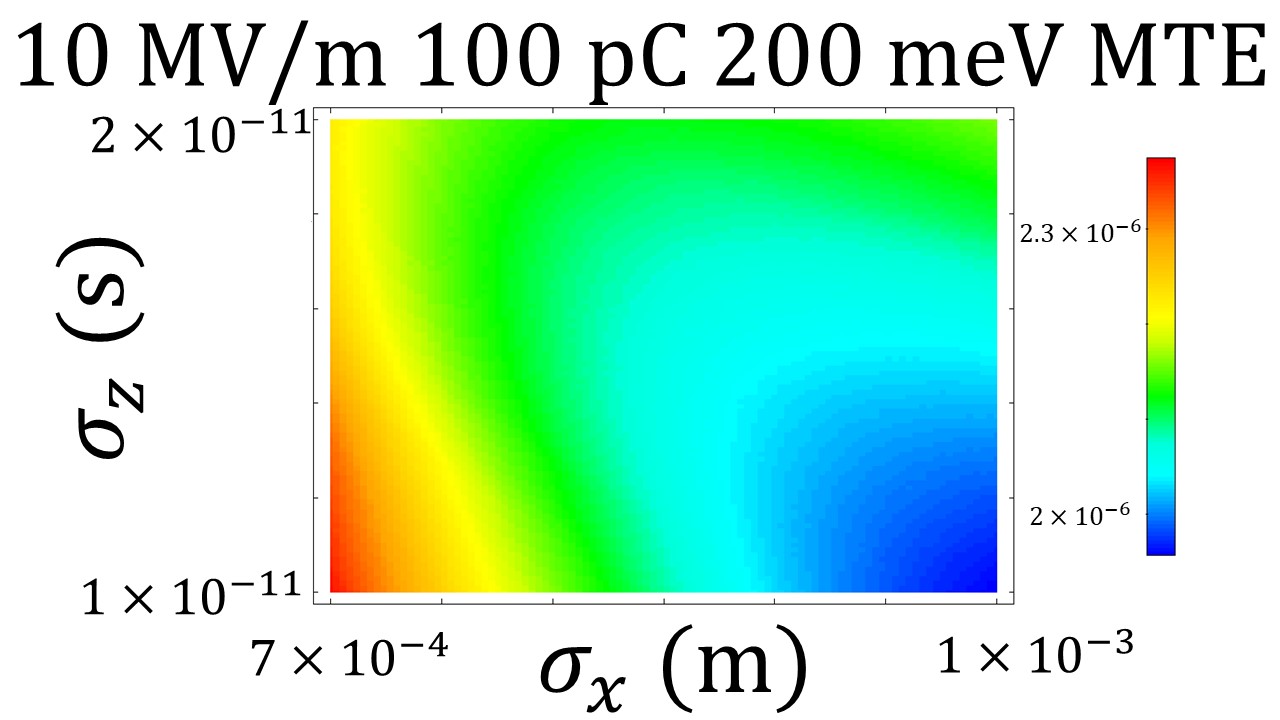}
	\includegraphics[width=0.45\textwidth, height=0.245\textwidth]{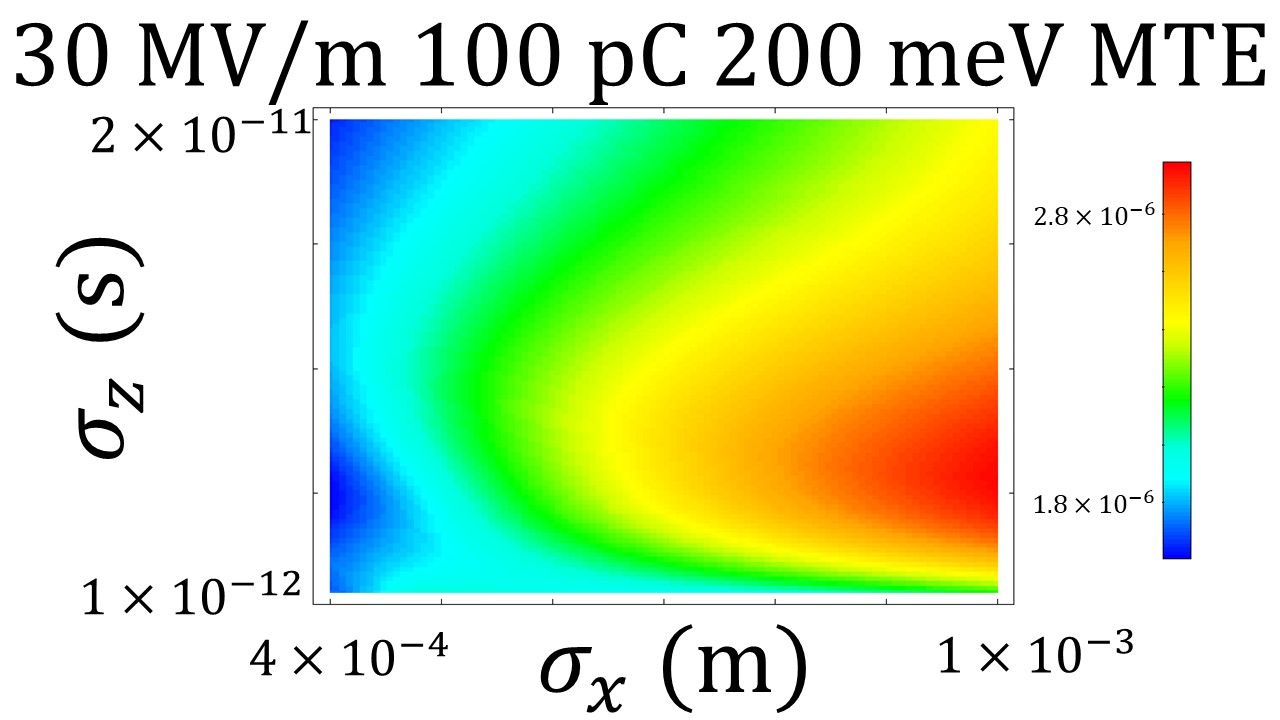}
	\includegraphics[width=0.45\textwidth, height=0.245\textwidth]{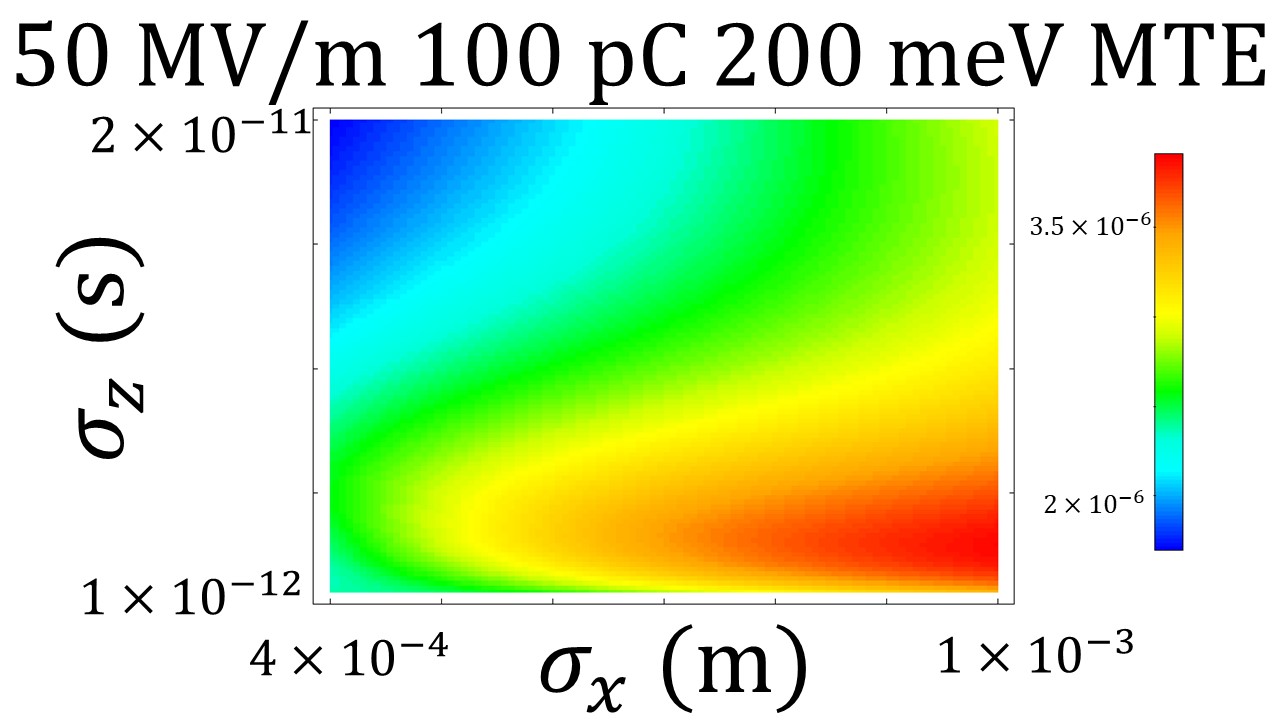}
	\includegraphics[width=0.45\textwidth, height=0.245\textwidth]{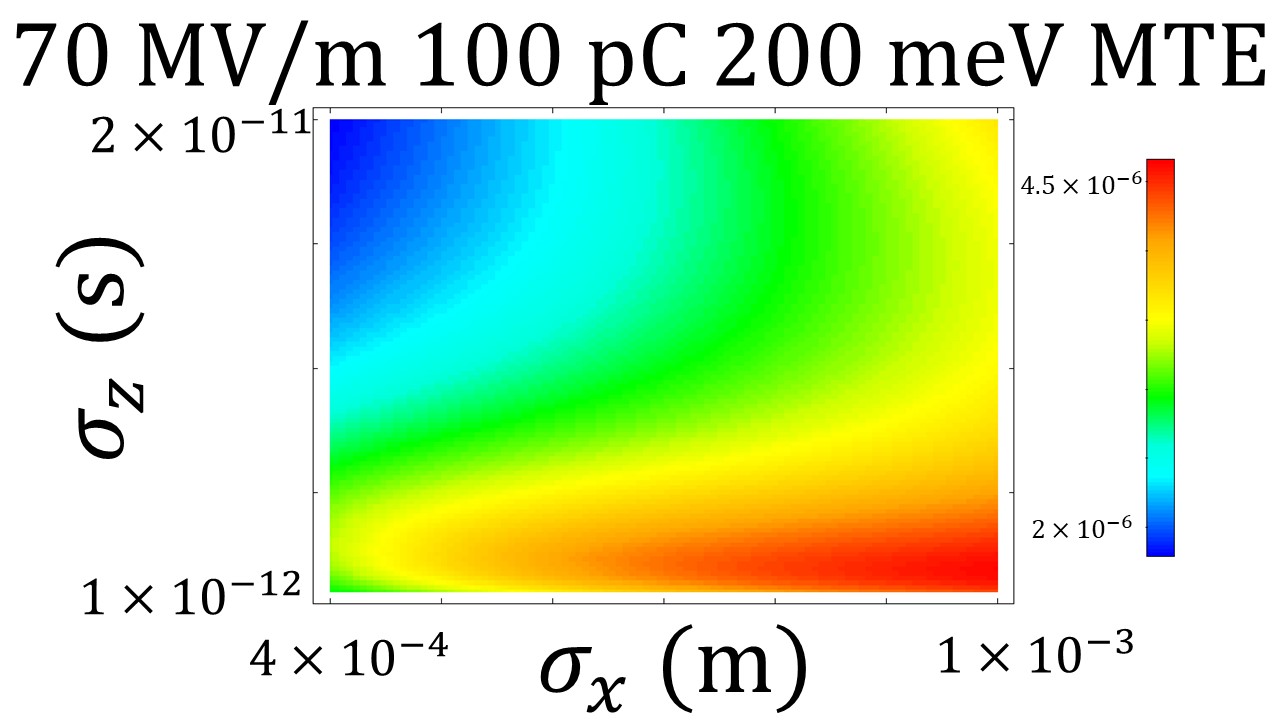}
	\includegraphics[width=0.45\textwidth, height=0.245\textwidth]{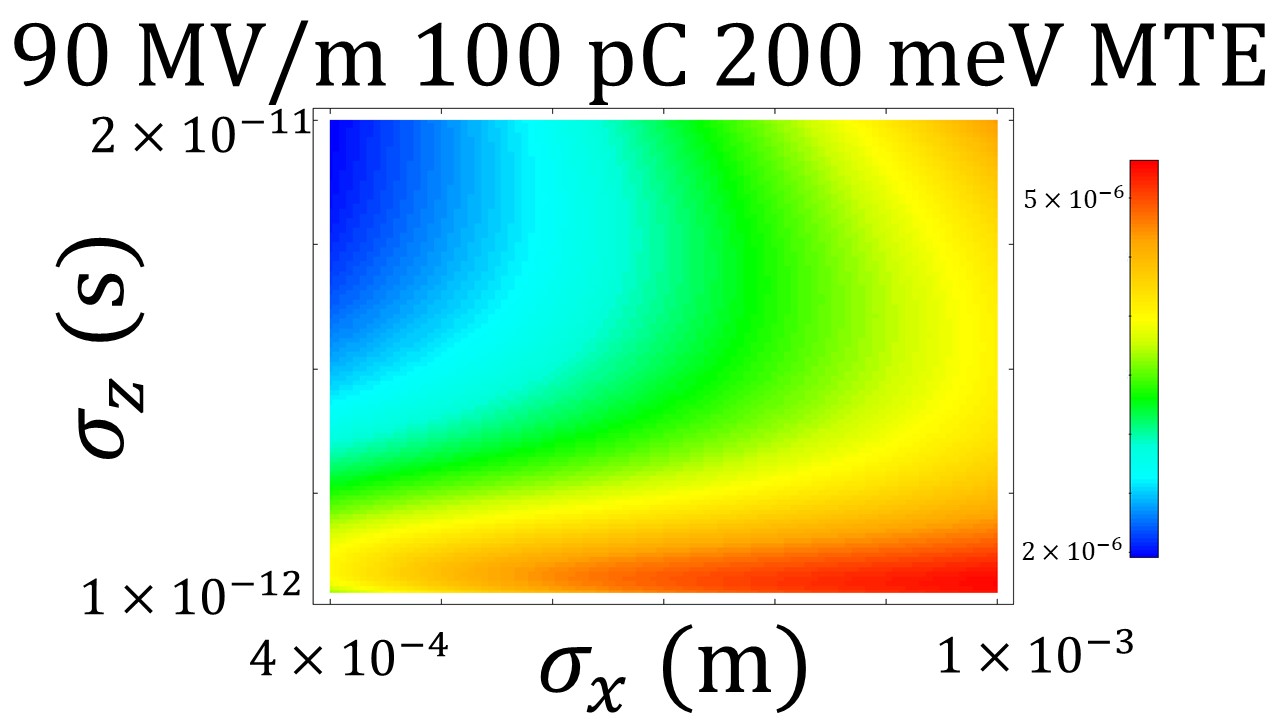}
\caption{Series of heatmaps for 100 pC and 200 meV MTE case study demonstrating beam form factor evolution at different gradients. Inclusion of intrinsic emittance leads to minor increase in emittance across the board and a reduction in minimal emittance area as compared to Fig.\ref{Fig 5}.}
\label{Fig 6}
\end{figure}

Compared to the 100 pC 0 MTE case study, the new set of the heatmaps remained nearly the same as can be seen in Fig.\ref{Fig 6}. The space charge effect is exceptionally strong, and therefore conceals the intrinsic emittance effect, despite the fact that $\sigma_x$ has to be the largest possible. The only effect the intrinsic emittance has, is that it slightly increases the minimal emittance baseline from $1.5\times 10^{-6}$ to $2\times 10^{-6}$ m.

\section{Discussion}\label{Discus}
There were two effects observed:

\noindent 1) The interactions between the space charge and intrinsic emittance were immediately understood and discussed in Section \ref{GPT}. In short, minimizing space charge is best attained with a large cross section bunch while minimizing intrinsic emittance is best accomplished with a tight small cross section bunch. This results in two counter processes and an optimized solution should be found to minimize both emittances such that total emittance is minimal.

\noindent 2) General trends suggesting that for best practices (lowest emittance and thus highest brightness) the bunch (regardless of the bunch charge and intrinsic emittance) has to be reshaped from the pancake aspect ratio, as under low gradient, to the cigar aspect ratio, as under high gradient. This result merits a separate discussion. To do that, the individual emittance components are analyzed, as the total emittance can then be calculated through the summation of the individual emittance components. The individual emittance components considered are the rf emittance, space charge emittance, and the intrinsic emittance, as defined in Eq.\ref{eq:8}, which is set to zero as it fundamentally does not affect the pancake to cigar switching.

The space charge and rf emittance components can be written as \cite{KIM1989201}

\begin{gather}
   \varepsilon_{sc} = \frac{\pi I}{4 \alpha k I_A \sin{\theta}}\frac{1}{3\frac{\sigma_x}{\sigma_z} + 5}\ \label{eq:9} \\
   \varepsilon_{rf} = \frac{\alpha k^3 \sigma_x^2 \sigma_z^2 }{\sqrt{2}} \label{eq:10}
\end{gather}

where \(\alpha=\frac{e E}{2 m k c^2}\). It is important to note that the \(\sigma_z\) and \(\sigma_x\) described here are not the initial pulse length and spot sizes, but the \(\sigma_z\) and \(\sigma_x\) of the bunch as it travels through the beam line. As the space charge emittance is proportional to the emitted current, the current is required. The current is conventionally charge per time. However, the time component depends on the shape of the bunch. To resolve this convoluted situation, we analyzed two distinct case. For a long bunch, where $\sigma_z$ is larger than $\sigma_x$, the time is defined by Eq.\ref{eq:11} derived in Ref.\cite{Filippetto}. The long pulse length, in this case, allows the electron ample time to escape the surface and is not a limiting factor.
 \begin{gather}
   \delta t_{Pulse} = \sqrt{\frac{2 m \sigma_z}{e E}} \cong \sqrt{\frac{\sigma_z}{E}}\label{eq:11}
\end{gather}
 In the contrasting case, where $\sigma_x$ is larger than $\sigma_z$, the dynamics occurring at the surface require a different definition of the characteristic time scale. As the pulse length is short in this case, there is a characteristic minimum time for the electron to be emitted from the surface and can described as \cite{Filippetto}
  \begin{gather}
   \tau_{Responce} = \sqrt{\frac{2 m \sigma_x}{e E}} \cong \sqrt{\frac{\sigma_x}{E}} \label{eq:12}
\end{gather}

\begin{figure*}[]
    \includegraphics[width=0.4\textwidth, height=0.25\textwidth]{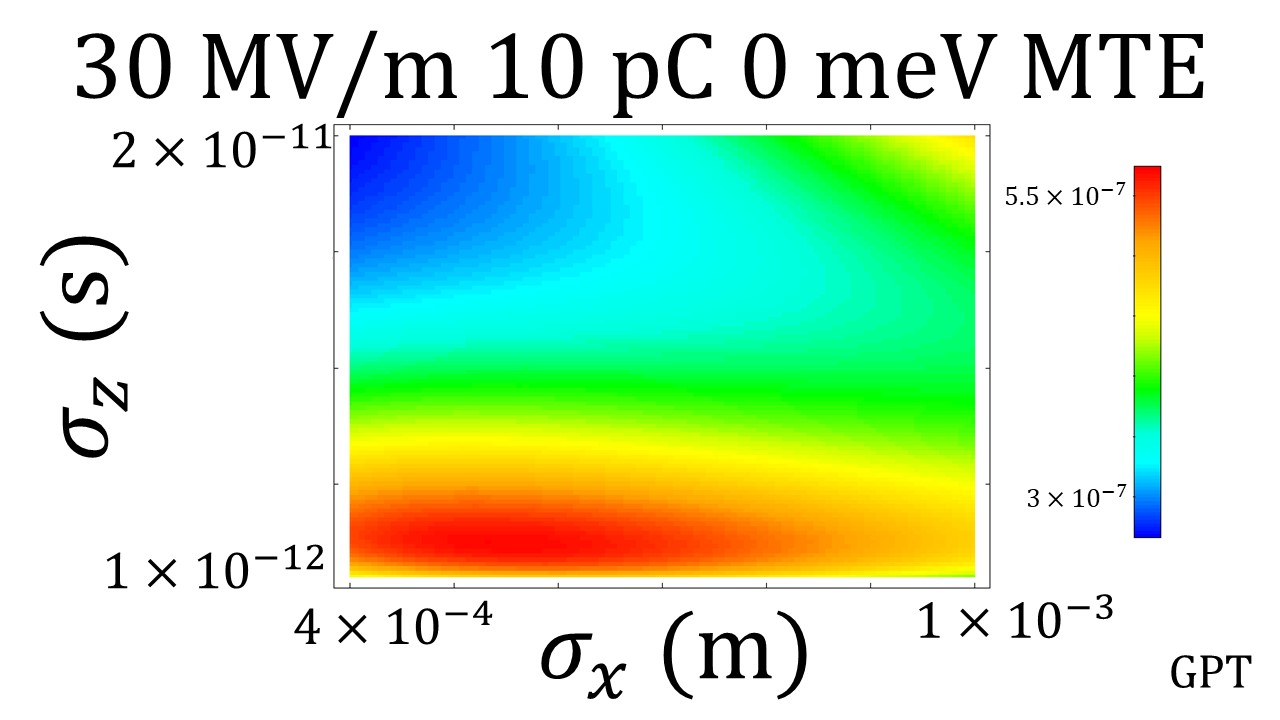}
	\includegraphics[width=0.4\textwidth, height=0.25\textwidth]{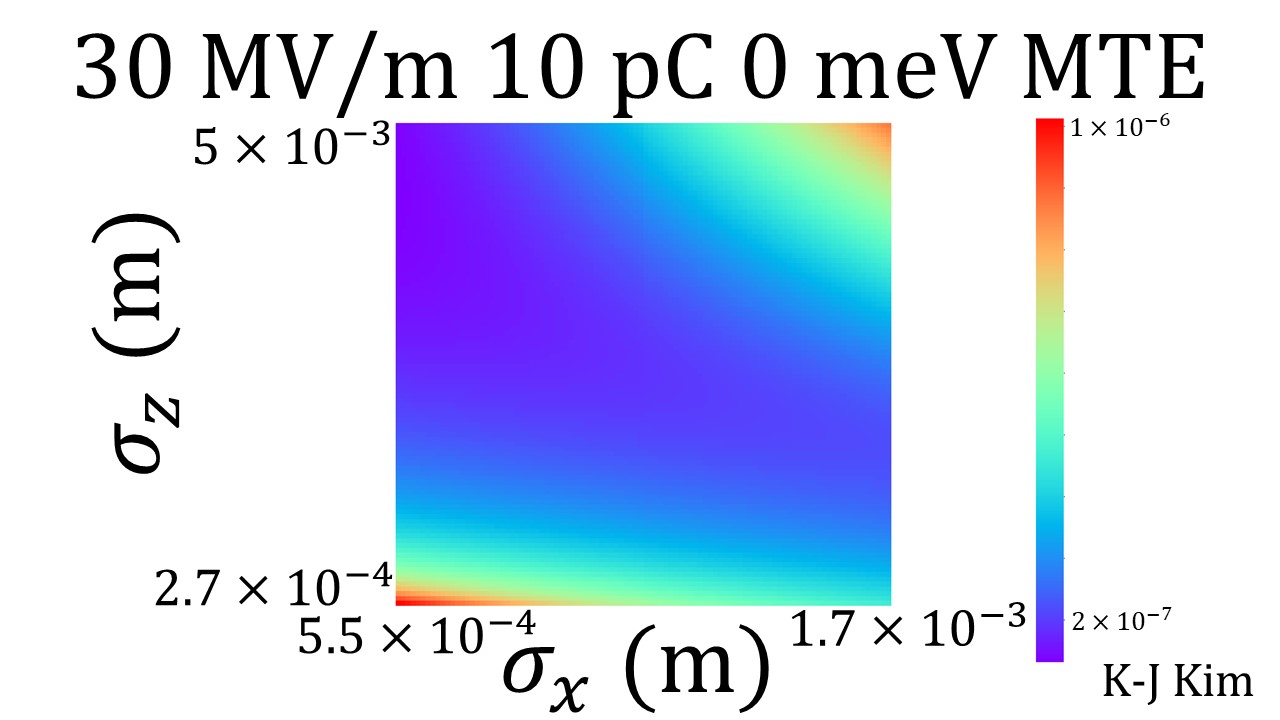}
	\includegraphics[width=0.4\textwidth, height=0.25\textwidth]{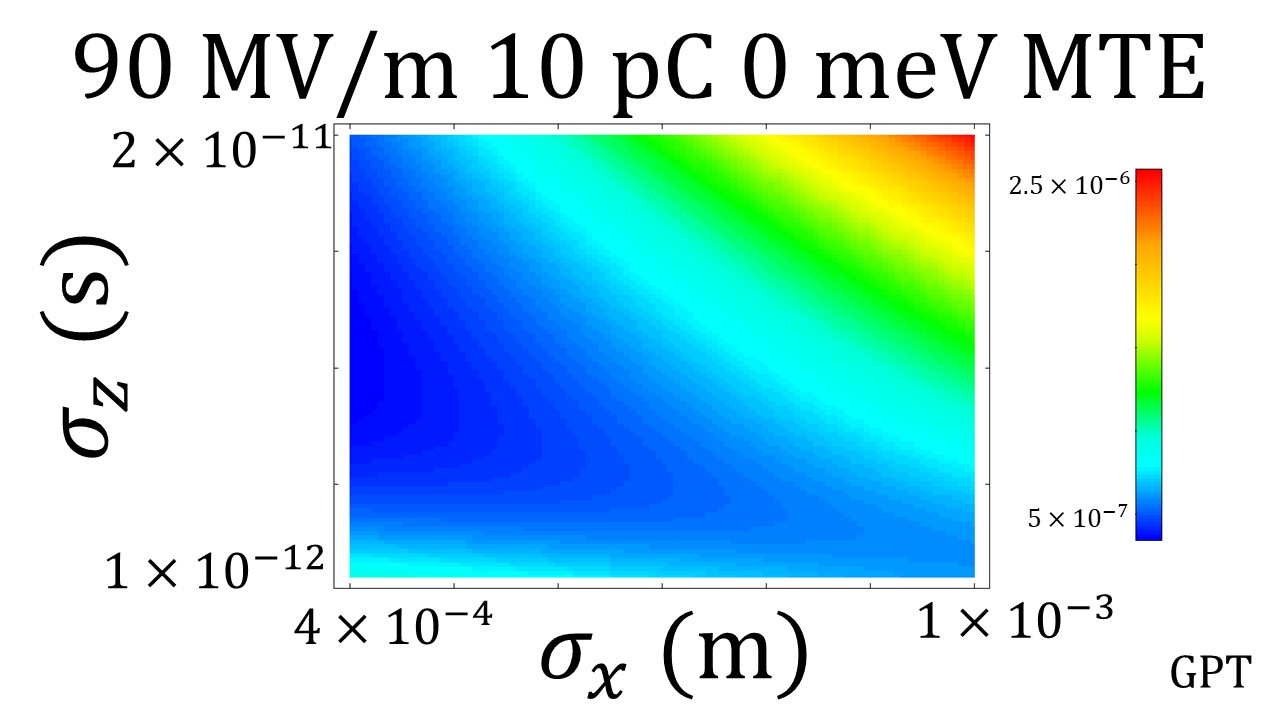}
	\includegraphics[width=0.4\textwidth, height=0.25\textwidth]{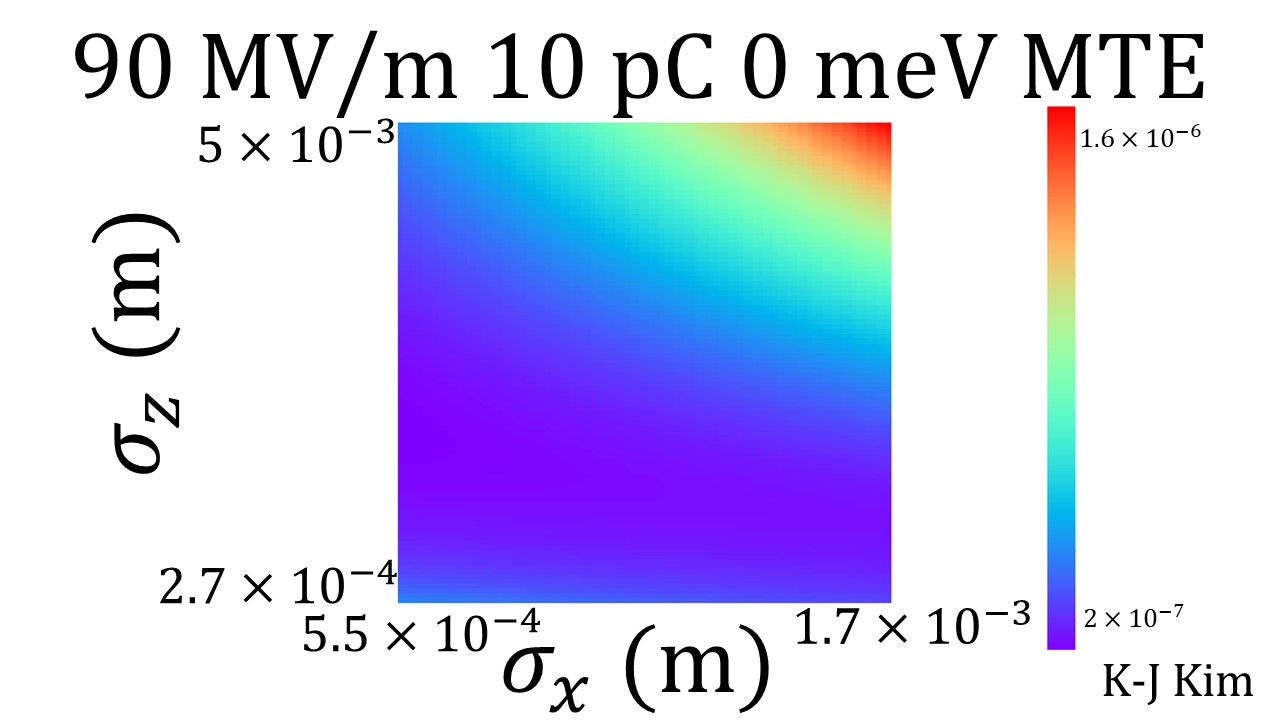}
\caption{Back-to-back comparison of the emittance heatmaps for 30 and 90 MV/m obtained using multivariate optimizer and analytical computation using K-J Kim’s emittance formulation. Bunch charge was 10 pC at 0 MTE.}
\label{Fig 7}
\end{figure*}
 These two cases, when linked with the emittance equations above, allow for analysis of the relations between the gradient, $\sigma_x$ and $\sigma_z$.
In the $\sigma_x < \sigma_z$ case, substitution of Eq.\ref{eq:11} to Eq.\ref{eq:9} yields Eq.\ref{eq:13}, which creates a situation where the only variables that can be changed to minimize emittance are the gradient and \(\sigma_z\). Both values would need to be increased to minimize the space charge emittance. Thus, the case of $\sigma_x < \sigma_z$, or the cigar beam case, is equivalent to the high gradient case, as the electric field needs to be increased to decrease the emittance. The rf emittance as in Eq.\ref{eq:14} derived from Eq.\ref{eq:10} would then increase in this situation, which is why the spot size \(\sigma_x\) needs to be reduced to compensate, thereby making the cigar aspect ratio more pronounced. Such analytical formalism therefore fully explains the consistency between the high gradient and the cigar beam regime as visualized by the obtained heatmaps.
\begin{gather}
   \varepsilon_{sc} = \frac{Q}{\sqrt{E}}\frac{1}{\sqrt{\sigma_z}} \label{eq:13} \\
   \varepsilon_{rf} \approx E \sigma_x^2  \sigma_z^2 \label{eq:14}
\end{gather}
For the $\sigma_x > \sigma_z$, or pancake aspect ratio, we define time as in Eq.\ref{eq:12}. In this case, the space charge emittance can be simplified as seen in Eq.\ref{eq:15}, while rf emittance as in Eq.\ref{eq:14}. It becomes clear that, in order to, first and foremost, minimize the space charge emittance, term $\sigma_z$ has to be decreased and $\sigma_x$ has to be increased. The growth of the rf emittance term associated with the $\sigma_x$ growth is countered by the drop in $\sigma_z$. Again, it fully explains the consistency between the low gradient and the pancake beam regime as visualized by the obtained heatmaps.
  \begin{gather}
    \varepsilon_{sc} = \frac{Q}{\sqrt{E}}\frac{\sigma_z}{\sigma_x^{\frac{3}{2}}}\label{eq:15} 
\end{gather}
 The results of the data are further confirmed when the $\sigma_x$ and $\sigma_z$ produced at the end of the ACT gun simulations were used in Eqs.\ref{eq:9} and Eq.\ref{eq:10} and the resulting emittance, mapped in $\sigma_z$ and $\sigma_x$ parameter space, was directly compared to K-J Kim's formalism where the total emittance is calculated as $\varepsilon = \sqrt{\varepsilon_{sc}^2 + \varepsilon_{rf}^2}$. The comparison results at 30 and 90 MV/m are presented in Fig.\ref{Fig 7}, both at 10 pC and 0 MTE. Analytical method produces similar heatmaps clarifying the nature of the computational results plotted in Figs.\ref{Fig 2}-\ref{Fig 6}.
 
 The main differences between the plots can be attributed to the dynamic change in the bunch dimensions as it travels along the injector. K-J Kim’s formalism uses the physical dimensions of the bunch, not the laser parameters, i.e. pulse length and spot size. Therefore, the boundaries of the K-J Kim’s plots had to be approximated using typical values of bunch dimensions that were acquired at the exit plane of the injector.

\section{Conclusion}\label{Conc}
Multivariate optimization of a quarter wave high gradient injector highlights fundamental relationships between the gradient and spatio-temporal bunch profile. Analytical treatment of the problem using classical emittance formulation and two-dimensional space charge model was able to capture the basic link between the bunch form factor and the injector gradient hence granting further insights to predict optimal parameters which can then be used to inform design decisions for injectors and laser systems. Future injector design can thus benefit from the presented parametrization that enables general understanding of the fundamental trade-offs. For instance, high brightness, high charge, and small energy spread cannot conveniently co-exist, but by designing the injector for a preferred application it can help negate interactions between those processes. By choosing the desired application (single shot microscopy with high charge versus spectroscopy with small energy spread) and then utilizing those requirements to influence injector design it can allow for minimizing complexity of the beamline optical element lattice.

\section{Acknowledgments}\label{Acknow}
The work by Benjamin Sims was supported by the U.S. Department of Energy Office of Science, High Energy Physics under Cooperative Agreement Award No. DE-SC0018362. The work by Sergey Baryshev was supported by the U.S. Department of Energy, Office of Science, Office of High Energy Physics under Award No. DE-SC0020429. We are grateful to Dr. John Lewellen for his insightful discussions, helpful advice, and ingenious sequencer. 

\bibliography{references}

\appendix
\addcontentsline{toc}{section}{Appendices}
\renewcommand{\thesubsection}{\Alph{subsection}}

\end{document}